\documentclass[conference,letterpaper]{IEEEtran}

\usepackage{cite}
\usepackage{graphicx}
\usepackage[cmex10]{amsmath} 
\usepackage{amssymb,amsfonts}
\usepackage{pgfplots}
\usepackage{float}
\usepackage{verbatim}
\usepackage{textcomp}
\usepackage{xcolor}

\DeclareMathOperator*{\argmin}{\arg\!\min}
\DeclareMathOperator*{\argmax}{\arg\!\max}

\usepackage[utf8]{inputenc} 
\usepackage[T1]{fontenc}
\usepackage{url}
\usepackage{ifthen}

\usepackage{mathtools}
\usepackage{dsfont}
\mathtoolsset{showonlyrefs=true}
\usepackage[noend]{algpseudocode}

\usepackage{color}

\newcommand{\crc}{{\rm crc}}
\newcommand{\aug}{{\rm aug}}
\newcommand{\thr}{{\rm thr}}

\newcommand{\ch}{{\rm ch}}

\newcommand{\rem}[1]{}

\newcommand{\bre}{\begin{equation}}
\newcommand{\ere}{\end{equation}}
\newcommand{\be}{{\bf {e}}}
\newcommand{\ee}\]
\newcommand{\bra}{\begin{eqnarray}}
\newcommand{\era}{\end{eqnarray}}
\newcommand{\bfg}{\begin{figure}[hbtp]}
\newcommand{\efg}{\end{figure}}
\newcommand{\bver}{\begin{verbatim}}
\newcommand{\ever}{\end{verbatim}}
\newcommand{\bit}{\begin{itemize}}
\newcommand{\eit}{\end{itemize}}
\newcommand{\ben}{\begin{enumerate}}
\newcommand{\een}{\end{enumerate}}

\newcommand{\bg}{{\bf{g}}}
\newcommand{\bG}{{\bf{G}}}

\newcommand{\bF}{{\bf F}}

\newcommand{\bll}{\boldsymbol{\ell}}

\newcommand\norm[1]{\lVert#1\rVert}

\newcommand{\baa}{\begin{eqnarray*}}
\newcommand{\eaa}{\end{eqnarray*}}

\newcommand{\bs}{{\bf s}}

\newcommand{\bH}{{\bf H}}

\newcommand{\xor}{\oplus}
\newcommand{\bu}{{\bf u}}
\newcommand{\bv}{{\bf v}}

\newcommand{\bx}{{\bf x}}
\newcommand{\by}{{\bf y}}
\newcommand{\bz}{{\bf z}}

\newcommand{\cA}{{\cal A}}
\newcommand{\cP}{{\cal P}}

\newcommand{\defined}{\triangleq}

\def\argmax{\mathop{\rm argmax}}
\def\argmin{\mathop{\rm argmin}}

\def\defined{\: {\stackrel{\scriptscriptstyle \Delta}{=}} \: }

\newfont{\boldlarge}{msbm10 scaled 1100}

\newlength{\tmpbigbar}

\begin{document}
\title{Belief Propagation List Ordered Statistics Decoding of Polar Codes}
\author{%
	\IEEEauthorblockN{Guy Mogilevsky and David Burshtein}
	\IEEEauthorblockA{School of Electrical Engineering\\
		Tel-Aviv University\\
		Tel-Aviv 6997801, Israel\\
		Email: guym1@mail.tau.ac.il, burstyn@eng.tau.ac.il}
}

\maketitle

\begin{abstract}
It is shown how to combine ordered statistics decoding (OSD) with CRC-aided belief propagation list (CBPL) decoding of polar codes. Even when the reprocessing order of the OSD is as low as one, the new decoder is shown to significantly improve on CBPL.
For reprocessing orders higher than one, we suggest partial reprocessing, where only error patterns associated with the least reliable part of the belief propagation decoded most reliable independent bits are considered. This type of partial reprocessing offers a trade-off between performance and computational complexity.
\end{abstract}


\section{Introduction} \label{sec:Introduction}

The performance of polar codes \cite{PolarArikan2009} under successive cancellation (SC) decoding can be significantly improved by concatenating the polar code with a high rate cyclic redundancy check (CRC) code and using CRC-aided successive cancellation list (CA-SCL) decoding \cite{SCLTal2015}.
However, the SCL decoder suffers from high decoding latency due to its serial nature. Various modifications and improvements to SC and SCL have been suggested to address this problem, e.g. \cite{alamdar2011simplified,leroux2013semi,sarkis2014fast,li2014low,balatsoukas2015llr,yuan2015low,xiong2015symbol,chen2016reduce,hashemi2018decoder,hashemi2018decoding,giard2018fast,hashemi2019rate}.

In \cite{PolarBPArikan} it was shown that belief propagation (BP) decoding over the code's factor graph (FG) can be used to improve upon the SC decoder. Efficient implementations, improvements and extensions of BP decoding for polar codes were suggested in \cite{BP_arc,BP_BEC,PolarBPConcat,bp_early_term,PolarBPTermination,PolarBPCRCWarren,BPPermuted,BPL,BPPermutedWarren,PolarBPCRCBrink,wang2019belief,ranasinghe2020partially,ren2020efficient,yu2019belief}. 
In particular, it was suggested to implement a BP list (BPL) decoder \cite{BPL} that applies BP decoding on different permutations of the polar FG layers \cite{PolarSource2009} in parallel.
It was also suggested to apply BP decoding over a concatenation of the CRC and polar FGs \cite{PolarBPCRCWarren, PolarBPCRCBrink}.
Yet, even when combining these ideas, the resulting CRC-aided BPL (CBPL) decoder \cite{PolarBPCRCBrink} has a higher error rate compared to CA-SCL.

Another family of decoders for polar codes are based on ordered statistics decoding (OSD) \cite{OSD95}. This approach was used in \cite{OSDPolar16, OSDPolar19B}, in which OSD
was either applied directly on the channel output, or combined with CA-SCL or SC.
Ordered statistics decoding and its box and match variant were also proposed in \cite{trifonov2011generalized,trifonov2012efficient,xu2017distance,OSDPolar19C,goldin2019performance} for polar and polar-like codes when viewed as a generalized concatenated code \cite{blokh1974coding}.
However, the required reprocessing order is typically large. Even though there are known methods for improving the complexity-performance trade-off of OSD \cite{valembois2004box, jin2006reliability, jin2006enhanced, wu2006soft, jin2006probabilistic, wu2007soft}, the computational complexity may still be prohibitive.

In the context of OSD decoding of low-density parity-check (LDPC) codes it was suggested \cite{OSDBP2001} to combine BP and OSD decoding by using the soft decoded BP output to rank the code bits rather than using the uncoded information as in plain OSD.
In this paper, we start by adapting the approach in \cite{OSDBP2001} to CBPL decoding of polar codes. We apply OSD reprocessing of order one using the soft decoding output of each of the parallel CBP decoders in CBPL.
Our simulations show that even OSD with reprocessing order as low as one can significantly decrease the error rate of CBPL, and bring it closer to that of CA-SCL for a relatively small permutation list size.
For reprocessing orders higher than one, we suggest partial reprocessing, where only error patterns associated with the least reliable part of the belief propagation decoded most reliable independent bits are considered. This type of partial reprocessing offers a trade-off between performance and computational complexity.
We demonstrate that partial order-2 reprocessing, incorporating only the least reliable half (or even less) of the total number of pairs of most reliable independent bits, results in an error-rate similar to that of full reprocessing.

\section{Background} \label{sec:backg}
\subsection{CRC augmented polar codes} \label{sec:backg_crc_aug}
Consider the $N \times N$ binary matrix:
\begin{equation}
\bG_N= \bF^{\otimes n} \label{eq:GN}
\end{equation}
where $\bF^{\otimes n}$ is the $n$-fold Kronecker product of the \textit{standard polarization kernel}
$\bF = \begin{bmatrix} 1 & 0 \\ 1 & 1 \end{bmatrix}$ and $n=\log_2{N}$.
Denote by $\cA$ the length $K$ \textit{information set} of a length $N$ polar code, $\cP(N,K)$. Then  $\cA \subseteq \{1,2,\ldots,N\}$.
The \emph{frozen set} is $\cA^\mathsf{c} \defined \{1,2,\ldots,N\}\setminus\cA$.
Denote the matrix composed of the rows of $\bG_N$, corresponding to the elements of $\cA$, by $\bG_N(\cA)$. The polar code, $\cP(N,K)$, is a linear binary block code with generator matrix $\bG_N(\cA)$ and code dimension $K$.
Encoding a message vector $\bu_{\cA}$ of length $K$ with the generator matrix $\bG_N(\cA)$ (using $\bu_{\cA} \bG_N(\cA)$) is equivalent to encoding the vector $\bu$ with generator matrix $\bG_N$ (using $\bu \bG_N$), where $\bu$ is a vector of length $N$ with message bits $\bu_{\cA}$ and with $\bu_{\cA^\mathsf{c}} = \mathbf{0}$ (i.e., we assume that the frozen bits are all zeros).

Also consider a CRC code which appends $r$ CRC bits to binary vectors of length $m$. This code can be represented as a linear binary block code with code rate $R_{\crc} = m / (m+r)$, a systematic generator matrix $\bG_{\crc}$ of dimensions $m \times (m+r)$, and a parity-check matrix (PCM) $\bH_{\crc}$ with $r$ parity check constraints.

For the polar code and CRC code described above, the corresponding \textit{CRC-augmented polar code} is the linear binary block code of length $N$, dimension $m$ and generator matrix $\bG_{\aug} = \bG_{\crc} \bG_N(\cA)$. Note that we set $K=m+r$, and that the overall code rate is $R = m / N$.

In our setup, a message word of length $m$ is encoded into a codeword $\bx=(x_1,\ldots,x_{N})$ of length $N$ using $\bG_{\aug}$, which is then BPSK-modulated and sent across a binary input additive white Gaussian noise channel (BIAWGNC) with noise $Z \sim N(0,N_0 / 2)$, so that the channel output is given by,
\begin{equation}
\by = \text{BPSK}(\bx) + \bz \label{eq:BIAWGNC}
\end{equation}
where $\text{BPSK}(\bx)\triangleq (-1)^{\bx} \triangleq((-1)^{x_1},\cdots,(-1)^{x_{N}})$. The log likelihood ratios (LLRs) of the codeword's bits, $\ell^{\ch}_i$, based on the respective channel output, $y_i$, are given by
\begin{equation}
\ell^{\ch}_i =
\ln \frac{\Pr(X_i=0 \mid y_i)}{\Pr(X_i=1 \mid y_i)} = 2\frac{y_i}{\sigma^2}
\label{eq:LLRxi}
\end{equation}
for $i=1,\ldots,N$, where $\sigma^2 = N_0 / 2$.

\subsection{CRC-aided belief propogation list decoding} \label{sec:backg_crc_bpl}
\begin{figure}[htbp]
\centerline{\includegraphics[width=0.9\columnwidth]{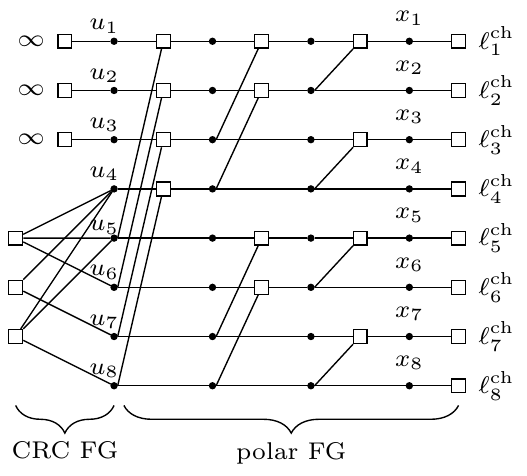}}
\caption{FG of a CRC-augmented polar code of length $N=8$, rate $R=1/4$ (2 message bits) and a 3-bit CRC.
The frozen set is $\cA^\mathsf{c}=(u_1,u_2,u_3)$.
}
\label{fig:PolarFG}
\end{figure}

In \cite{PolarBPArikan} it was suggested to apply BP decoding on the FG representing the encoding with the generator matrix in \eqref{eq:GN}. This FG is shown in Fig. \ref{fig:PolarFG}. The BP decoder can be implemented in parallel, thus reducing the latency and increasing the throughput of the decoder compared to SC and SCL decoding.
The BP uses left and right-propagating messages over the edges of the FG. Denote by $L^t_{i,l}$ ($R^t_{i,l}$, respectively) the left-propagating (right-propagating) message out of (into) variable node $i$ in the $l$'th FG layer at iteration $t$, for $0\leq l\leq n$ and $1\leq i\leq N$. An iteration of BP consists of a right-to-left message propagation, in which messages are sent from the right-most layer of the FG to the left-most layer, followed by a similar left-to-right propagation of messages. The left-most (right-most, respectively) layer represents the LLRs of the message and frozen bits (codeword bits).
The initialization of the messages (at iteration $t=0$) is,
\begin{equation}
L^0_{i,n} = \ell^{\ch}_i,
\qquad
R^0_{i,0}=\begin{cases}
0,      & \text{$i\in \cA$}\\
\infty, & \text{$i\in \cA^\mathsf{c}$}
\end{cases}
\end{equation}
for $1\leq i\leq N$.
The detailed iterative message passing equations are provided in \cite{PolarBPArikan}.

To reduce the error-rate, it was suggested to
perform the message passing on the concatenation of the CRC FG, represented by the PCM $\bH_{\crc}$, and the polar FG. An example is shown in Fig. \ref{fig:PolarFG}. As stated in \cite{PolarBPCRCWarren}, for this concatenation to be beneficial in terms of error-rate performance, the messages must first evolve through $I_{\thr}$ iterations on the polar FG alone, so that the information layer LLRs will become reliable enough.
The BP iterations are performed until either a predetermined maximum number of iterations $I_{\max}$ have been reached, or until a certain stopping condition has been met.
Following \cite{bp_early_term,PolarBPTermination}, in our implementation we stop iterating if the following two conditions are satisfied. The first condition is
$\hat{\bx}=\hat{\bu}\cdot\bG_N$
where $\hat{\bu}$ and $\hat{\bx}$ are the hard-decisions of the information LLRs and codeword LLRs, respectively.
The second condition is that $\hat{\bu}_{\cA}$ satisfies the CRC constraints ($\hat{\bu}_{\cA}\bH_{\crc}^T=\mathbf{0}$).

To further reduce the error-rate, we can perform CBP on a list of $L$ layer-permuted polar FGs \cite{PolarSource2009,BPPermuted,BPL}, thus obtaining the CRC-aided BP List (CBPL) decoder \cite{PolarBPCRCBrink}.
The estimated codeword is the CBP output $\hat{\bx}$ with BPSK representation closest, in Euclidean distance, to the channel output, $\by$, out of all the outputs that are valid codewords (namely, the outputs of the CBP instances in which the stopping condition was met).
To simplify the implementation, instead of using different polar FGs in the CBP realizations, we may permute their inputs and outputs \cite{BPPermutedWarren}.

\subsection{Ordered statistics decoding} \label{sec:backg_osd}
Ordered statistics decoding (OSD) \cite{OSD95} is a general method for decoding a linear binary block code.
Consider a linear binary block code of blocklength $N$ and $K$ information bits transmitted over the BIAWGNC \eqref{eq:BIAWGNC}. The code is represented by a $K\times N$ full row rank generator matrix, $\bG$.
The OSD algorithm consists of two main parts \cite{OSD95}: Finding the most reliable independent basis (MRIB) from the columns of $\bG$ with respect to the values in $\bll=\bll^{\ch}$ (computed from the channel output $\by$ according to \eqref{eq:LLRxi}), and a reprocessing stage.
The MRIB is a set of $K$ columns from $\bG$, which correspond to the indices in $\bll$ that contain the most reliable LLRs (LLRs with highest absolute values) under the constraint that these columns are linearly independent (over GF(2)). The process of finding the MRIB is described in detail in \cite{OSD95}.
It starts by sorting the absolute values of the components of $\bll$ in decreasing order. We then apply the same ordering (permutation) on the columns of $\bG$. Following that, we use Gaussian elimination conducted on the above mentioned permuted $\bG$ in order to find the first $K$ positions of linearly independent columns, which serve as the MRIB. In the end of the process, we have a new matrix, $\tilde{\bG}$, which is an equivalent representation of the code up to some permutation, $\lambda$, of the code bits. The first $K$ columns of $\tilde{\bG}$ are the MRIB columns of $\bG$ with respect to $\bll$.
The above mentioned Gaussian elimination is implemented such that $\tilde{\bG}$ is represented in a systematic form, i.e.
$$
\tilde{\bG}=[\ \textbf{I}_K\ |\ \textbf{A}\ ]
$$
where $\textbf{I}_K$ is the $K\times K$ identity matrix, and $\textbf{A}$ is a $K\times (N-K)$ matrix.
Denote by $\tilde{\bll}=\lambda(\bll)$ the permutation of the LLRs vector, $\bll$, using the same permutation $\lambda$.
The first $K$ values in $\tilde{\bll}$ (which are the most reliable LLRs) can be used to obtain an initial estimate to the 
information vector $\hat{\bv}=(\hat{v}_1,\cdots,\hat{v}_{K})$ (corresponding to the systematic generator matrix $\tilde{\bG}$), by using hard decisions:
\begin{equation}
\hat{v}_i=\begin{cases}
0, & \text{$\tilde{\ell}_{i}\geq 0$}\\
1, & \text{else}
\end{cases}
\label{eq:hvi}
\end{equation}
for $1\leq i\leq K$.
We can now start the reprocessing stage of the OSD algorithm.
For each $0\leq i \leq q$, flip all possible combinations of $i$ bits in $\hat{\bv}$, i.e., consider $\hat{\bv}\xor\be$ ($\xor$ is the bit-wise XOR operator) for all error patterns $\be$ of Hamming weight at most $q$ (the total number of error patterns is $\sum\limits_{i=0}^q \binom{K}{i}$). For each error pattern, $\be$, re-encode $\hat{\bv}\xor\be$ using $\tilde{\bG}$, and calculate the Euclidean distance between the BPSK representation of the resulting codeword and the permuted channel output vector, $\tilde{\by} = \lambda(\by)$. Keep track of the distances, so that after all possible error patterns of Hamming weight at most $q$ have been tested, select the (permuted) codeword with minimum distance from $\tilde{\by}$, and inversely permute it by $\lambda^{-1}$ to obtain the OSD($q$) estimate of the transmitted codeword.

\section{Inclusion of OSD in CBPL decoding} \label{sec:osd_cbpl}
In order to implement OSD decoding efficiently for LDPC codes, it was proposed \cite{OSDBP2001} to determine $\bll$ from the soft output of a preliminary BP decoder, rather than using $\bll=\bll^{\ch}$. This is usually advantageous, since it will make the LLRs in $\bll$ more reliable (more LLRs will have the correct sign). 

\begin{figure}[htbp]
\centerline{\includegraphics[width=\columnwidth]{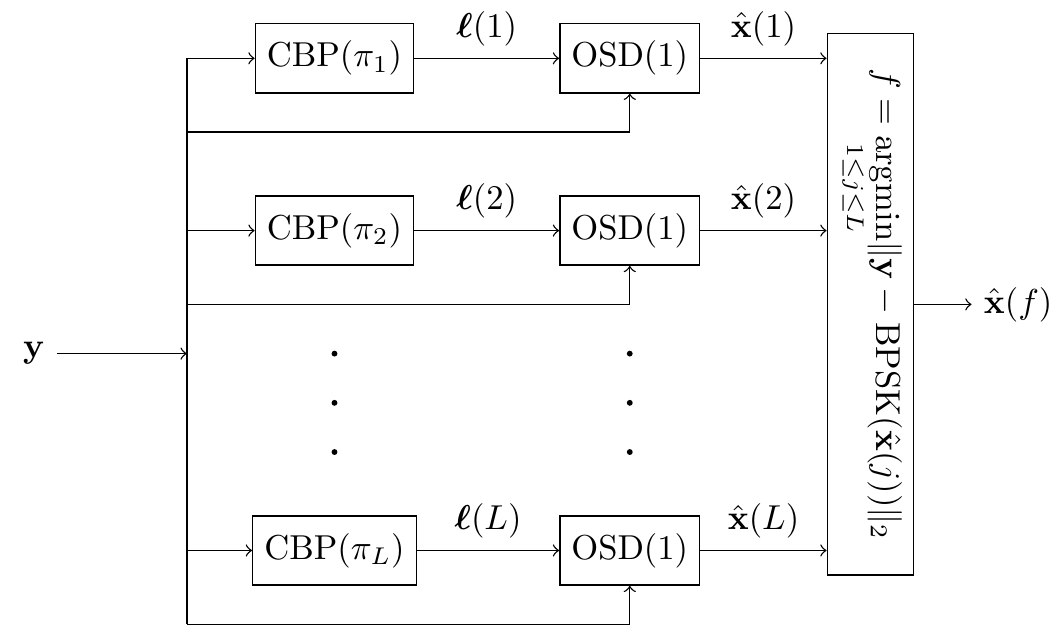}}
\caption{The CBPLOSD(1) scheme. The permutation set is $S=\{\pi_j\}^L_{j=1}$, and CBP$(\pi_j)$ denotes CBP with a polar FG layer permutation $\pi_j$.}
\label{fig:CBPLOSD1}
\end{figure}

We assume the setup described in Section \ref{sec:backg}, where a codeword $\bx$ from a CRC-augmented polar code of length $N=2^n$ and with generator matrix $\bG_{\aug}$ is transmitted over the BIAWGNC, resulting in a channel output $\by$.
Our suggestion for further improving the error performance of CBPL decoding is to perform OSD$(q)$ on the soft output of each of the $L$ CBP instances running in parallel, as depicted in Fig. \ref{fig:CBPLOSD1} for $q=1$.
We denote this decoder with reprocessing order $1$ by CBPLOSD($1$).
Interestingly, as we report later on, even for $q=1$ we obtain a substantial improvement on the performance of the CBPL algorithm.
On the other hand, plain OSD decoding requires much higher values of $q$ to reach the same performance, with a much higher total number of operations.
A further improvement is obtained by using CBPLOSD(2).
This is considered in Section \ref{sec:pool}.

In Fig. \ref{fig:CBPLOSD1}, the channel output $\by$ is the input to $L$ parallel CBP decoders with different polar FG layer permutations.
After meeting the stopping criterion (from section \ref{sec:backg_crc_bpl}) or completing the maximum number of iterations, $I_{\max}$, in the $j$-th CBP decoder, $j=1,\ldots,L$, we apply OSD($1$) decoding using its soft output LLR vector, $\bll(j)$, which contains the decoded LLRs of the codeword bits.
The OSD(1) routine at the $j$'th branch, uses this soft output, along with $\by$ and $\bG_{\aug}$ as input, and outputs a codeword estimate $\hat{\bx}(j)$.
Note that an output of OSD is always a valid codeword. Thus, we simply choose the codeword estimate, $\hat{\bx}(f)$, with BPSK representation closest to $\by$ as the final output of the decoder,
\begin{equation}
f = \argmin_{1\leq j \leq L}\norm{\by-\mathrm{BPSK}(\hat{\bx}(j))}_2 \: .
\end{equation}

The reprocessing part of the OSD(1) decoder can be implemented efficiently as follows.
Denote by $\tilde{\bG}_{\aug}$ the permuted systematic generator matrix obtained from $\bG_{\aug}$ in the first stage of OSD described above (finding the MRIB). 
Denote
$$
\hat{\bv}^0 \defined \hat{\bv}
$$
where $\hat{\bv}$ is obtained from the LLRs of the bits corresponding to the MRIB as in \eqref{eq:hvi}.
Also,
$$
\hat{\bv}^{i} \defined \hat{\bv}\xor\be_i \quad i=1,\ldots,K
$$
is the information vector after flipping the $i$'th bit in $\hat{\bv}$. The corresponding (permuted) codeword is
$$
\hat{\bx}^{i} \defined \hat{\bv}^{i}\cdot\tilde{\bG}_{\aug} \: .
$$
Denote by $\bg^0 \defined (0,\ldots,0)$, and by $\bg^i$, the $i$'th row of $\tilde{\bG}_{\aug}$, $i=1,\ldots,K$. In order to calculate $\hat{\bx}^i$ for each $0\leq i\leq K$ efficiently, we can use the relation
$$
\hat{\bx}^i=\bar{\bx}\xor\bg^i
$$
where 
$$
\bar{\bx}=\hat{\bv}\cdot\tilde{\bG}_{\aug} \: .
$$
Furthermore, it can be easily verified that
\begin{equation}
i_0 =
\argmin_{0\leq i\leq K}\norm{\tilde{\by}-\mathrm{BPSK}(\hat{\bx}^i)}_2 =
\argmax_{0\leq i\leq K}\sum_{l=1}^{N}s_l\cdot(-1)^{g^i_l}
\label{eq:minL2}
\end{equation}
where
$$
\textbf{s}=(\tilde{y}_1\cdot(-1)^{\bar{x}_1},\ldots,\tilde{y}_{N}\cdot(-1)^{\bar{x}_{N}}) \: .
$$
The decoder outputs $\lambda^{-1}(\hat{\bx}^{i_0})$. Using the right-hand-side (RHS) of \eqref{eq:minL2}, the reprocessing, which can be implemented in parallel, requires $(N-1)(K+1)$ additions / subtractions.

The Gaussian elimination over GF(2) required in the first stage of OSD can also be implemented in parallel in hardware \cite{FastGausElim}.

\section{Partial higher order OSD reprocessing} \label{sec:pool}
So far we have focused on CBPLOSD(1), as it has the lowest computational complexity. Yet, a lower error rate may be achieved by using CBPLOSD($q$) for $q>1$. In this section, we propose approximating higher order reprocessing by performing it only on the bits that are associated with the least reliable LLRs. We demonstrate this approximation for CBPLOSD(2), and show that it can be used to achieve almost the same performance as that of regular CBPLOSD(2), while cutting back half or more of the reprocessing complexity.

Recall the information vector $\hat{\bv}$ from \eqref{eq:hvi} and the systematic $K\times N$ generator matrix $\tilde{\bG}=\tilde{\bG}_{\aug}$, where $\bg^i$ denotes its $i$'th row. Also recall the vector $\textbf{s}$ from \eqref{eq:minL2}. Let us further denote by $\hat{\bv}^{(i,j)}$ the resulting information vector after flipping the $i$'th and $j$'th bits in $\hat{\bv}$, and its corresponding codeword by
$$
\hat{\bx}^{(i,j)}=\hat{\bv}^{(i,j)}\cdot\tilde{\bG}_{\aug} \: .
$$
The main bottleneck of order-2 reprocessing is the search for the codeword $\hat{\bx}^{(i,j)}$ with BPSK closest to the permuted channel output $\tilde{\by}$ in terms of Euclidean distance,
\begin{equation}
\argmin_{1\leq i<j\leq K}\norm{\tilde{\by}-\mathrm{BPSK}(\hat{\bx}^{(i,j)})}_2 \: .
\label{eq:minL2A}
\end{equation}
To determine the final OSD(2) estimate, the result of the search \eqref{eq:minL2A} needs to be compared to the OSD(1) estimate in \eqref{eq:minL2} by minimum Euclidean distance to $\tilde{\by}$.

In \eqref{eq:minL2A} we search over $\binom{K}{2}=[K(K-1)] / 2$ pairs of indices $(i,j)$. Similarly to \eqref{eq:minL2}, we have the following relation,
\begin{align}
\lefteqn{\argmin_{1\leq i<j\leq K}\norm{\tilde{\by}-\mathrm{BPSK}(\hat{\bx}^{(i,j)})}_2 =}\\
&&
\argmax_{1\leq i<j\leq K}\sum_{l=1}^{N} s_l\cdot(-1)^{g^i_l}\cdot(-1)^{g^j_l} \: .
\label{eq:minL2B}
\end{align}
Using the RHS of \eqref{eq:minL2B}, the complexity of this search, and thus of order-2 reprocessing, is about $N K^2 / 2$ additions.

The RHS of \eqref{eq:minL2B} can be expressed using matrix multiplication (that only requires additions / subtractions, and can be implemented efficiently in parallel) as follows.
First define a $K\times N$ matrix $A$ such that the $i$'th row, $i=1,\ldots,K$, of $A$ is $\bs \cdot (-1)^{\bg^i}$ (element-wise multiplication and exponentiation). Next, define an $N \times K$ matrix $B$ such that the $j$'th column, $j=1,\ldots,K$, of $B$ is $(-1)^{\bg^j}$.
Finally define the $K\times K$ matrix $C$ by
\begin{equation}
C = A \cdot B \: .
\end{equation}
Then the RHS of \eqref{eq:minL2B} can be written as
\begin{equation}
\argmax_{1 \le i < j \le K} c_{i,j} \: .
\end{equation}
A similar formulation using matrix multiplication applies to the partial reprocessing method that we now suggest.

Recall the permuted LLRs $(\tilde{\ell}_1,\ldots,\tilde{\ell}_K)$, corresponding to the MRIB, defined in Section \ref{sec:backg_osd}. To reduce the complexity we propose the following approximation to the search in \eqref{eq:minL2B}. For any integer $M < [K(K-1)] / 2$, we perform this search only on the $M$ pairs of indices $(i,j)$ with the lowest values of $|\tilde{\ell}_i|+|\tilde{\ell}_j|$. Namely, we flip the $M$ pairs of bits in $\hat{\bv}$ which are associated with the least reliable LLRs (LLRs with the lowest absolute values), and enumerate the $M$ resulting codewords using the RHS of \eqref{eq:minL2B}. Since $|\tilde{\ell}_i| > |\tilde{\ell}_j|$ for $1\leq i < j \leq K$, this is implemented by enumerating over the indices $(i,j)$ during reprocessing in the following decreasing order: $i=K-1,K-2,\ldots$ in the outer loop and $j=K,K-1,\ldots,i+1$ in the inner loop, until we have exhausted $M$ such pairs. This approximation can be straightforwardly extended for higher orders. In the case of order 2, it reduces the complexity of the search over the pairs from about $N K^2 / 2$ to about $N M$.

\section{Simulation Results} \label{sec:sim}
In this section we present decoding results for two CRC-augmented polar codes of rate $R=0.5$ with blocklengths $N=256$ and $N=512$.
The CRC generator polynomial for both codes is the 5G standardized 6-bit polynomial $g_6(x)=x^6+x^5+1$.
All the CBP decoders were set to use $I_{\max}=100$, $I_{\thr}=50$, and a total of $L=6$ such decoders for CBPL were used, with their input / layer permutations corresponding to all the $3!=6$ possible permutations of the $3$ right-most layers of the polar FG. The CA-SCL list size was also 6.

\subsection{CBPOSD(1) and CBPLOSD(1)} \label{sec:sim_cbplosd}
In Fig. \ref{fig:CBPLOSD1_N256} we compare the performances of plain OSD, CBP, CBPOSD(1), CBPL, CBPLOSD(1) and CA-SCL for the code with blocklength $N=256$. In Fig. \ref{fig:CBPLOSD1_N512} we repeat this comparison for the other code with $N=512$.

\definecolor{carminered}{rgb}{1.0, 0.0, 0.22}
\definecolor{ferngreen}{rgb}{0.31, 0.47, 0.26}
\begin{figure}[h]
	\centering
	\begin{tikzpicture}
	\begin{semilogyaxis}[
	legend columns=2, 
	legend pos = south west,
	xlabel={$E_b / N_0$ [dB]},
	ylabel={FER},
	xmin=1.5, xmax=3.5,
	xtick={1.5, 2.0, 2.5, 3.0, 3.5},
	ytick={},
	ymajorgrids=true,
	xmajorgrids=true,
	grid style=dashed,
	legend style={font=\small}
	]

	\addplot[  
	color=ferngreen,
	mark=triangle,
	]
	coordinates {
		(1.5, 10^-0.43)(2.0, 10^-0.70)(2.5, 10^-1.12)(3.0, 10^-1.54)(3.5, 10^-2.18)
	};

	\addplot[  
	color=orange,
	mark=square,
	]
	coordinates {
		(1.5, 10^-0.46)(2.0, 10^-0.83)(2.5, 10^-1.40)(3.0, 10^-2.04)(3.5, 10^-2.80)
	};

	\addplot[  
	color=brown,
	mark=x,
	]
	coordinates {
		(1.5, 10^-0.75)(2.0, 10^-1.18)(2.5, 10^-1.76)(3.0, 10^-2.45)(3.5, 10^-3.19)
	};	
	
	\addplot[  
	color=blue,
	mark=+,
	]
	coordinates {
		(1.5, 10^-0.60)(2.0, 10^-1.10)(2.5, 10^-1.76)(3.0, 10^-2.53)(3.5, 10^-3.31)
	};
	
	\addplot[  
	color=purple,
	mark=o,
	]
	coordinates {
		(1.5, 10^-1.05)(2.0, 10^-1.64)(2.5, 10^-2.38)(3.0, 10^-3.19)(3.5, 10^-3.98)
	};
	
	\addplot[  
	color=black,
	mark=-,
	]
	coordinates {
		(1.5, 10^-0.95)(2.0, 10^-1.60)(2.5, 10^-2.43)(3.0, 10^-3.47)(3.5, 10^-4.68)
	};
	
	\legend{OSD$(2)$ \\ 
		CBP \\
		CBPOSD(1) \\
		CBPL \\
		CBPLOSD(1) \\
		CA-SCL \\}
	\end{semilogyaxis}
	\end{tikzpicture}
	\caption{FER comparisons for a CRC-augmented polar code of length $N=256$ and rate $R=0.5$.}
	\label{fig:CBPLOSD1_N256}
\end{figure}
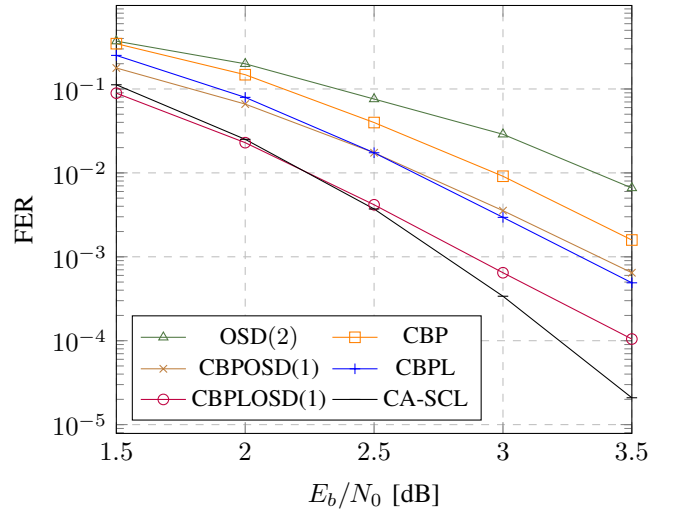
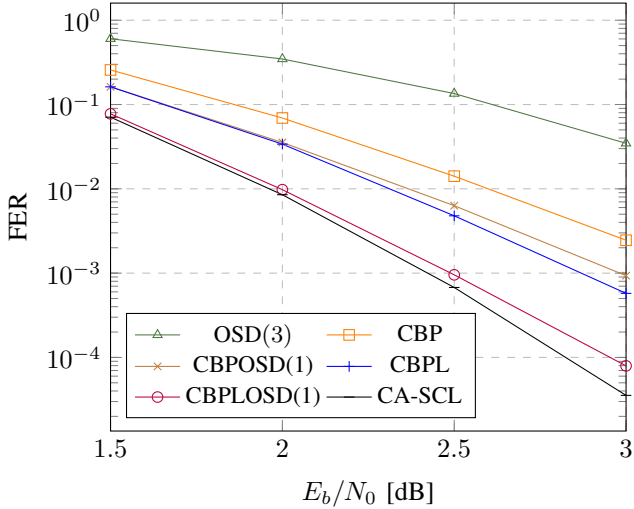
\begin{figure}[h]
	\centering
	\begin{tikzpicture}
	\begin{semilogyaxis}[legend columns=2, legend pos = south west,
	xlabel={$E_b / N_0$ [dB]},
	ylabel={FER},
	xmin=1.5, xmax=3.0,
	xtick={1.5, 2.0, 2.5, 3.0},
	ytick={},
	ymajorgrids=true,
	xmajorgrids=true,
	grid style=dashed,
	legend style={font=\small}
	]
	
	\addplot[  
	color=ferngreen,
	mark=triangle,
	]
	coordinates {
		(1.5, 10^-0.22)(2.0, 10^-0.46)(2.5, 10^-0.87)(3.0, 10^-1.46)
	};

	\addplot[  
	color=orange,
	mark=square,
	]
	coordinates {
		(1.5, 10^-0.59)(2.0, 10^-1.16)(2.5, 10^-1.85)(3.0, 10^-2.61)
	};

	\addplot[  
	color=brown,
	mark=x,
	]
	coordinates {
		(1.5, 10^-0.79)(2.0, 10^-1.45)(2.5, 10^-2.20)(3.0, 10^-3.03)
	};	
	
	\addplot[  
	color=blue,
	mark=+,
	]
	coordinates {
		(1.5, 10^-0.79)(2.0, 10^-1.47)(2.5, 10^-2.32)(3.0, 10^-3.24)
	};
	
	\addplot[  
	color=purple,
	mark=o,
	]
	coordinates {
		(1.5, 10^-1.11)(2.0, 10^-2.01)(2.5, 10^-3.02)(3.0, 10^-4.10)
	};
	
	\addplot[  
	color=black,
	mark=-,
	]
	coordinates {
		(1.5, 10^-1.15)(2.0, 10^-2.07)(2.5, 10^-3.17)(3.0, 10^-4.45)
	};
	\legend{OSD$(3)$ \\ 
		CBP \\
		CBPOSD(1) \\
		CBPL \\
		CBPLOSD(1) \\
		CA-SCL \\}
	\end{semilogyaxis}
	\end{tikzpicture}
	\caption{Same as Fig. \ref{fig:CBPLOSD1_N256} for blocklength $N=512$.}
	\label{fig:CBPLOSD1_N512}
\end{figure}
As can be seen, in the high SNR range, the gain in $E_b / N_0$ attributed to the inclusion of OSD$(1)$ in CBPL is around $0.5$dB for the $N=256$ code and $0.4$dB for the $N=512$ code.
When no permutations are used ($L=1$), incorporating OSD$(1)$ in CBP provides a $0.3$dB gain in the high SNR range for both codes.
Both figures demonstrate that even for a relatively small permutation list size ($L=6$), the performance of CBPOSD(1) comes close to that of CA-SCL with a similar list size. On the other hand, plain OSD requires a much larger reprocessing order to match the performances of CBPLOSD(1) and CA-SCL.

\subsection{Partial order-2 reprocessing} \label{sec:sim_partial}
In Fig. \ref{fig:PCBPLOSD256} we compare plain CBPLOSD(2), in terms of error performance, to CBPLOSD decoders in which the search in \eqref{eq:minL2B} was conducted only on a portion of the least reliable bits, using the CRC-augmented polar code with blocklength $N=256$.
We refer to the CBPLOSD decoder in which the partial order 2 reprocessing takes into account only $\alpha\cdot\binom{RN}{2}$ of the least reliable pairs $(0<\alpha<1)$ by P-CBPLOSD$(2,\alpha)$.

\definecolor{darkgreen}{rgb}{0.0, 0.2, 0.13}
\definecolor{darkcoral}{rgb}{0.8, 0.36, 0.27}
\definecolor{darktangerine}{rgb}{1.0, 0.66, 0.07}
\definecolor{auburn}{rgb}{0.43, 0.21, 0.1}
\definecolor{burntorange}{rgb}{0.8, 0.33, 0.0}
\definecolor{cadmiumred}{rgb}{0.89, 0.0, 0.13}
\definecolor{chocolate(traditional)}{rgb}{0.48, 0.25, 0.0}
\definecolor{cottoncandy}{rgb}{1.0, 0.74, 0.85}
\begin{figure}
	\centering
	\begin{tikzpicture}
	\begin{semilogyaxis}[
	legend columns=1, 
	legend pos = south west,
	xlabel={$E_b / N_0$ [dB]},
	ylabel={FER},
	xmin=1.5, xmax=3.0,
	xtick={1.5, 2.0, 2.5, 3.0},
	ytick={},
	ymajorgrids=true,
	xmajorgrids=true,
	grid style=dashed,
	legend style={font=\small}
	]
	\addplot[
	color=purple,
	mark=o,
	]
	coordinates {
		(1.5, 10^-1.05)(2.0, 10^-1.64)(2.5, 10^-2.38)(3.0, 10^-3.19)
	};
	\addlegendentry{CBPLOSD(1)}
	\addplot[
	color=blue,
	mark=x,
	]
	coordinates {
		(1.5, 10^-1.20)(2.0, 10^-1.86)(2.5, 10^-2.62)(3.0, 10^-3.39)
	};
	\addlegendentry{P-CBPLOSD$(2,\frac{1}{8})$}
	\addplot[
	color=chocolate(traditional),
	mark=+,
	]
	coordinates {
		(1.5, 10^-1.24)(2.0, 10^-1.91)(2.5, 10^-2.65)(3.0, 10^-3.42)
	};
	\addlegendentry{P-CBPLOSD$(2,\frac{1}{4})$}
	\addplot[
	color=cadmiumred,
	mark=-,
	]
	coordinates {
		(1.5, 10^-1.26)(2.0, 10^-1.95)(2.5, 10^-2.70)(3.0, 10^-3.47)
	};
	\addlegendentry{P-CBPLOSD$(2,\frac{1}{2})$}
	\addplot[
	color=darkgreen,
	mark=diamond,
	]
	coordinates {
		(1.5, 10^-1.27)(2.0, 10^-1.97)(2.5, 10^-2.76)(3.0, 10^-3.48)
	};
	\addlegendentry{CBPLOSD(2)}
	\end{semilogyaxis}
	\end{tikzpicture}
	\caption{FER comparisons between various CBPLOSD decoders, with full or partial order-2 reprocessing, for the CRC-augmented polar code of length $N=256$.}
	\label{fig:PCBPLOSD256}
\end{figure}
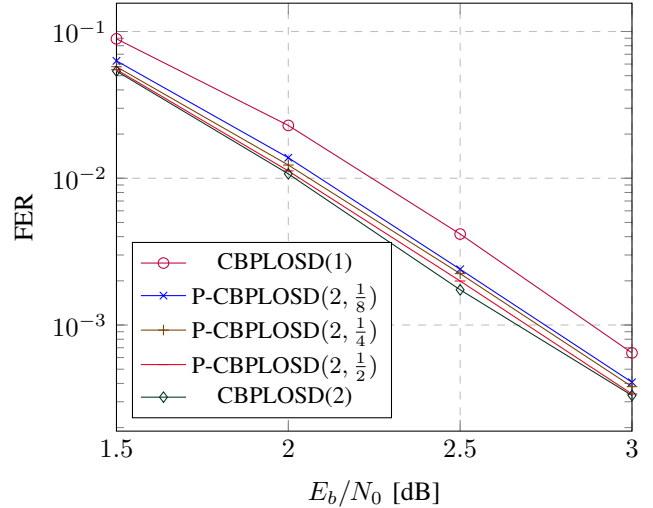
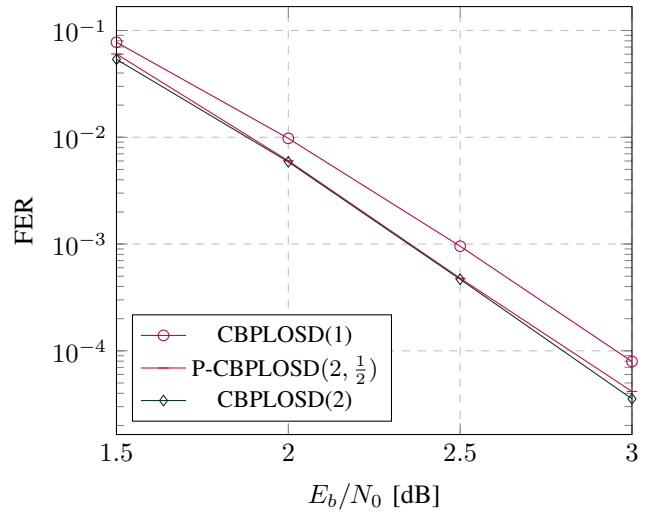
\begin{figure}
\centering
	\begin{tikzpicture}
	\begin{semilogyaxis}[
	legend columns=1, 
	legend pos = south west,
	xlabel={$E_b / N_0$ [dB]},
	ylabel={FER},
	xmin=1.5, xmax=3.0,
	xtick={1.5, 2.0, 2.5, 3.0},
	ytick={},
	ymajorgrids=true,
	xmajorgrids=true,
	grid style=dashed,
	legend style={font=\small}
	]
	\addplot[
	color=purple,
	mark=o,
	]
	coordinates {
		(1.5, 10^-1.11)(2.0, 10^-2.01)(2.5, 10^-3.02)(3.0, 10^-4.10)
	};
	\addlegendentry{CBPLOSD(1)}
	\addplot[
	color=cadmiumred,
	mark=-,
	]
	coordinates {
		(1.5, 10^-1.22)(2.0, 10^-2.22)(2.5, 10^-3.32)(3.0, 10^-4.38)
	};
	\addlegendentry{P-CBPLOSD$(2,\frac{1}{2})$}
	\addplot[
	color=darkgreen,
	mark=diamond,
	]
	coordinates {
		(1.5, 10^-1.27)(2.0, 10^-2.23)(2.5, 10^-3.33)(3.0, 10^-4.45)
	};
	\addlegendentry{CBPLOSD(2)}
	\end{semilogyaxis}
	\end{tikzpicture}
	\caption{Same as Fig. \ref{fig:PCBPLOSD256} for blocklength $N=512$.}
	\label{fig:PCBPLOSD512}
\end{figure}
We notice that CBPLOSD(2) shows a gain of about $0.2$-$0.3$dB in $E_b / N_0$ over CBPLOSD(1) in the simulated range. 
For the decoders with partial reprocessing, even when we go over only $1 / 8$ of the least reliable pairs (and, as a result, reduce the decoding complexity by about 8), the performance degradation, compared to full order-2 reprocessing, is about $0.1$dB or less in this example. This degradation can be mitigated by a small amount by going over $1 / 4$ or $1 / 2$ of the least reliable pairs instead, and the resulting performance comes close to that of full order-2 reprocessing.

Similar observations can be seen in Fig. \ref{fig:PCBPLOSD512} for the second code with $N=512$.

\section{Conclusions} \label{sec:conc}
We have shown how to combine OSD with CBPL decoding of polar codes. Even when the reprocessing order of the OSD was as low as one, the new decoder was shown to improve on CBPL in terms of the error performance by $0.4$-$0.5$dB in the high SNR region, for the two CRC-augmented polar codes considered.
Additional gain can be obtained by increasing the OSD reprocessing order to two.
Partial reprocessing, carried out only on the least reliable belief propagation decoded pairs of MRIB bits, can reduce the computational complexity of plain order-2 reprocessing by a significant amount, at the expense of a relatively small performance degradation.

\section*{Acknowledgment}
The authors would like to thank Yonatan Urman for his help with the simulation programs.
This research was supported by the Israel Science Foundation (grant no. 1868/18).

\clearpage



\end{document}